\documentclass[copyright,creativecommons]{eptcs}
\providecommand{\event}{HSB 2012} 
\usepackage{breakurl}             

\usepackage{graphicx}
\usepackage{graphics}
\usepackage{epsfig}
\usepackage{amsfonts}
\usepackage{amssymb}
\usepackage{amsmath}
\usepackage{placeins}
\usepackage{algorithmic}
\usepackage{algorithm}
\usepackage{sidecap}

\newcommand{\EE}{\mathbf{E}}
\newcommand{\Probab}{\mathcal{P}}

\title{Effects of delayed immune-response \\ in tumor immune-system interplay}
\author{Giulio Caravagna \quad Alex Graudenzi  \quad Marco Antoniotti  \quad Giancarlo Mauri
\institute{Dipartimento di Informatica Sistemistica e Comunicazione\\
Universit\`a degli Studi Milano-Bicocca,\thanks{G.C., A.G., G.M. and M.A. wish to acknowledge NEDD and the Regione Lombardia for financial support of this work, under the research project RetroNet, grant 12-4-5148000-40; U.A 053. }\\
 Viale Sarca 336, I-20126 Milan, Italy.}
\email{\{marco.antoniotti, giulio.caravagna, alex.graudenzi, mauri\}@disco.unimib.it}
\and
Alberto d'Onofrio 
\institute{Department of Experimental Oncology, \\
European Institute of Oncology, \\
Via Ripamonti 435, 20141 Milan, Italy.}
\email{alberto.donofrio@ieo.eu}
}
\def\titlerunning{Effects of a delayed immune-response in tumor immune-system interplay}
\def\authorrunning{G.Caravagna, A.Graudenzi,  A.d'Onofrio, M.Antoniotti \& G.Mauri}
\def\copyrightholders{Caravagna, Graudenzi, d'Onofrio, Antoniotti \& Mauri}
\begin{document}

%
%
%
%
%
%
%
%
%
%
%

\maketitle

\begin{abstract}
Tumors constitute a wide family of diseases kinetically characterized by the co-presence of multiple spatio-temporal  scales. So, tumor cells ecologically interplay with other kind of cells, e.g. endothelial cells or immune system effectors,  producing and exchanging various chemical signals. As such, tumor growth is an ideal object of hybrid modeling where discrete stochastic processes model  agents at low concentrations, and  mean-field equations model  chemical signals.   In previous works we proposed a hybrid version of the well-known Panetta-Kirschner mean-field model of tumor cells, effector cells and Interleukin-2. Our hybrid model suggested
-at variance of the inferences from its original 
formulation-  that immune surveillance, i.e. tumor elimination by the
immune system, may occur through a sort of side-effect of large stochastic
oscillations. However, that  model did not  account that,
due to both chemical transportation and  cellular differentiation/division,
the tumor-induced recruitment of  immune  effectors is not instantaneous but, instead, it exhibits a lag period. To capture this, we here integrate a mean-field equation for Interleukins-2 with a
bi-dimensional delayed stochastic process describing such delayed interplay. An algorithm to realize trajectories of the underlying stochastic process is obtained by coupling the Piecewise Deterministic Markov process  (for the hybrid part) with a Generalized Semi-Markovian clock structure  (to account for delays). We  $(i)$ relate tumor mass growth  with  delays via simulations and via parametric sensitivity analysis techniques, $(ii)$ we quantitatively determine probabilistic eradication times, and  $(iii)$ we prove, in the oscillatory regime, the existence of a heuristic stochastic bifurcation resulting in delay-induced tumor eradication, which is neither predicted by the mean-field nor by the hybrid non-delayed models.  
\end{abstract}

\section{Introduction}
Tumor--immune system interaction is triggered by the appearance of specific  antigens -- called neo-antigens -- eventually formed by the  vast number of genetic and epigenetic  events  characterizing tumors \cite{PA03}. So, the immune system may  control and, in some case to eliminate, tumors \cite{IS09}. This observation, fundamental to the so-called {\em immune surveillance} hypothesis,    recently accumulated evidences \cite{Dunn2004}.

The competitive interaction between tumor cells and the immune system is
extremely complex and, as such, it  has multiple outcomes. So, for instance,
a neoplasm may very often escape from immune control, may be constrained in a oscillatory regime or, differently, a dynamic equilibrium  with the tumor  in a microscopic undetectable ``dormant" 
steady-state \cite{CSFa} may also be established. In the oscillatory regime both  'short term-small amplitude' oscillations \cite{Kennedy,Vodo,RGatti,Mehta}  
and patterns of remission-recurrence \cite{PK36,PK3} have been observed, i.e. the alternation of long dormancy phases where the immune surveillance is not definitive with  tumor escape phases. The  latter case has important and negative implications since, on the one hand, a dormant tumor may eventually induce metastases through blood vessels formation and, on the other hand, the neoplasm may develop
strategies to circumvent the  immune system action, thus restarting to grow 
\cite{WH02,PA03,Dunn2004,VI02}. This evolutionary adaptation, termed ``immunoediting'',  typically happens over a significant fraction of the average host  life span \cite{Dunn2004} and, among its many effects, it negatively impacts on the
effectiveness of immunotherapies  \cite{donofrioPRE2011}. These therapies,  consisting in stimulating the immune system  to better fight, and hopefully eradicate, a cancer, are a simple and promising approach to the treatment of cancer \cite{[VHR]}, even though a huge inter-subjects variability is observed, which makes the results of
immunotherapy clinical trials quite puzzling \cite{[Seminar],[Euronc],[Kaminski]}.

As far as the modeling of  tumor--immune system interplay is concerned, many 
mean-field models have appeared  \cite{KP,[Kuznetsov],[KK],[dePillis],CSFa,donofrioPRE2011}, some of them including delays \cite{Todorovic,Villasana,DelayTumor}. However, since tumor cells  exchange a number of chemical signals with other kind of cells, e.g endothelial cells or  immune system effectors, they are an ideal object of hybrid modeling where some agents are represented by discrete stochastic processes, especially those in low numbers \cite{G77}, and chemicals are represented  by mean-field equations \cite{NoiJTB,NoiBMC}. This allows to consider the  {\em intrinsic noise} of the model and, when the mean-field approach would be an over-approximation, this may provide more  informative forecasts \cite{NoiJTB}. 

In \cite{NoiJTB,NoiBMC} we proposed a hybrid version of the well-known Panetta-Kirschner \cite{KP} mean-field model  of tumor cells, effector cells and Interleukins-2. 
The original model forecasts various kinds of experimentally observed tumor size
oscillations \cite{Kennedy,Vodo,RGatti,Mehta,PK36}, as well as 
microscopic/macroscopic  constant equilibria. However,  its hybrid analogous suggests -- in addiction to replicating  original deterministic forecasts --  that immune surveillance, i.e. tumor elimination by the
immune system, may occur through a sort of side-effect of large stochastic
oscillations. By discretizing both tumor and effector cellular populations, and by approximating the interleukins with a mean-field equation, probabilistic   tumor eradication times  s have been quantitatively determined  for various model configurations. Also, in \cite{NoiBMC}   the model was extended to account for both interleukin-based therapies and Adoptive Cellular Immunotherapies, i.e.   the transfusion of autologous or allogeneic T cells into tumor-bearing hosts \cite{ACI}, and  model outcomes have been investigated  under various therapeutic settings .

However, that hybrid model did not take into  account that,
due to both chemical transportation and  cellular differentiation/division,
the influence of tumor on immune system effectors recruitment and
proliferation is not instantaneous but, instead, it exhibits a lag period.
Thus, to represent this phenomenon, we here couple the mean-field equation for Interleukins-2 with a
bi-dimensional delayed stochastic process describing such a delayed interplay. This delay serves to approximate missing dynamical components, e.g. exchanged chemical signals, maturation and activation of T-lymphocytes mediated by B-lymphocytes \cite{Albe16} or, more in general, the fact that the immune system needs time to identify a tumor and react properly \cite{Albe40}. Of course, a full phenomenological model of these processes would be desirable. However, attempting to model each relevant stage of this process is currently impossible also because of the lack of systematic data \cite{Todorovic}.  Thus,  despite this abstraction being a highly macroscopical and simplistic representation of tumor--immune system interplay,  it can still provide useful insights in understanding this very fundamental and complex interaction.


This new hybrid system with delay is  a stochastic process  combining  the Piecewise Deterministic Markov process \cite{PDMP} underlying the delay-free model \cite{Bortolussi1,Bortolussi2,Bortolussi3} with a superimposed clock structure of  a Generalized Semi-Markov process \cite{GSMP1}, as one of those underlying chemically reacting systems with delays \cite{Iotesi,IoJane}. As a consequence,  numerical realizations of the model are obtained by combining a Gillespie-like Stochastic Simulation Algorithm with delays \cite{PDA} with the algorithm to simulate the delay-free hybrid system  \cite{NoiJTB}.  Via numerical analyses  $(i)$ we study the effect of various delays on tumor mass growth, $(ii)$ we quantitatively determine eradication times  as probability distributions, $(iii)$ we define a time-dependent sensitivity coefficient relating tumor mass and delay amplitude and  $(iv)$ we prove, in the oscillatory regime, the existence of a heuristic stochastic bifurcation resulting in delay-induced tumor eradication, which is neither predicted by the mean-field model nor by the hybrid non-delayed model.

The paper is structured as follows. In Section \ref{sec:model} we present the model
 with  delay, discuss its formulation in terms of hybrid automata and the underlying stochastic processes. In Section \ref{sec:analysis} we discuss algorithms for the realization of such  processes and, in  Section \ref{sec:results}, we present the results of our simulations. Finally, in Section \ref{sec:conclusions} we draw some conclusions and discuss future works.

\section{Model definition} \label{sec:model}

We start by extending the model given in \cite{NoiJTB,NoiBMC}  with the simple form of {\em constant delay} in the immune-response. We  consider two cell populations, i.e. tumor cells $T$ and   immune system effectors $E$, and the molecular population of Interleukins-2 (IL-2)  $I$. A {\em Delay Differential Equation} (DDE) model can be stated  by  considering two equations for cells
\begin{align} \label{eq:KPcells}
&T^{\prime}= r T \left(1- \frac{b}{V} T\right) - \frac{p_T T }{g_T V + T}E  &&E^{\prime}=\frac{p_E I}{g_E +I}E-\mu_E E + c T(t-\theta)
\end{align}
 and one equation for ILs-2, that is
\begin{eqnarray} \label{eq:KPil}
I^{\prime}= \frac{p_I}{V} \frac{ T E}{g_I V + T} -\mu_I I\, . 
\end{eqnarray}
These equations are obtained, as in \cite{NoiBMC}, by converting into total number of cells the densities $T_\ast$ and $E_\ast$ of the  mean-field model in \cite{KP} (not shown here), i.e. $T_\ast = T/V$ and $E_\ast={E}/{V}$ where  $V$ is the blood and bone marrow volumes for leukemia. In \cite{NoiJTB} an hybrid model is built by switching  to a  discrete representation of the populations ruled by equation  (\ref{eq:KPcells}) and by keeping continuous IL-2, as we shall discuss in the following. An immediate consequence of this is that equation  (\ref{eq:KPcells})  is interpreted  as a set of stochastic events, whereas equation (\ref{eq:KPil}) is left unchanged.
In this model the tumor induces the recruitment of the effectors at a linear rate $c T(t-\theta)$ with delay $\theta \geq 0$. With respect  to \cite{NoiBMC}, where instead the recruitment is instantaneous, i.e. $\theta=0$, the delay effect  is  to approximate missing dynamical components \cite{Albe16,Albe40}.
As in the original model formulation  $c$ is a measure of the immunogenicity of the tumor, i.e.  $c$ is  ``a measure of how different the tumor is from self'' \cite{KP}. Biologically, $c$  corresponds to the average number of antigens, i.e. secreted antibodies and/or surface receptors on immune system T-cells, expressed by each tumor cell.
 Interleukins stimulate effectors proliferation, whose  average lifespan  is $\mu_E^{-1}$, and   the average degradation time for IL-2 is $\mu_I^{-1}$. The source of interleukin is modeled as  depending on both the effectors and  the tumor burden. Michaelis-Menten kinetics rule IL-2 production by the tumor immune-system interplay, effectors recruitment by their interplay with IL-2 and effectors-induced tumour death. Finally, tumor growth is logistic with plateau $1/b$ .

In \cite{NoiJTB} it is shown that, when  $\theta=0$, the hybrid model predicts a desired {\em tumor  eradication via immune surveillance}, whereas the mean-field analogous does not \cite{KP}.  Subsequently, in \cite{NoiBMC}  
Adoptive Cellular Immunotherapies  and Interleukin-based therapies are added to the model. By focusing on realistic therapeutic settings, i.e. impulsive and piece-wise constant infusion delivery schedule, it is shown that  the delivery schedule   deeply impacts on the therapy-induced tumor eradication time. The advantage of resetting the mean-field version model to the hybrid setting allows to quantitatively determine the probability of  eradication, i.e.  $\Probab[T(t)=0]$ for some $t$,  given various model configurations.

In  hybrid systems terminology,  when $\theta=0$ this model  is a {\em Stochastic Hybrid Automaton} (SHA, \cite{Bortolussi1,Bortolussi2}) with  modes  in $\mathbb{N} \times \mathbb{N}$, i.e. the ``control'' part of the automaton, recording the cellular concentrations. 
The SHA consists of a mode for each possible value of $E$ and $T$,  i.e. a mode $q=(q_E,q_T)$ to count $q_e$ and $q_T$ effector and tumor cells, with inside the vector field of equation (\ref{eq:KPil}), i.e. such a mode  contains 
\begin{align}\label{eq:Imode}
I(t)= B_q + \left(I_q-B_q\right) \exp\left(-\mu_I(t-t_q)\right)
\end{align}
with initial condition $I(t_q)=I_q$ when $t_q$ is the mode entrance time  and $B_q =[p_Iq_T q_E/(g_I V^2 + q_TV)]/\mu_I$.
An automata execution switches probabilistically between modes,  while  continuous paths of  $I(t)$ are determined; so, 
when jumping from mode $q$, at time $t_q$, to  mode $q'$, at time $t_{q'}$, the initial condition of $I(t)$, i.e.  $I(t_{q'})$, is set equal to the last evaluation of $I(t)$, i.e. $I(t_{q})$. Jumps between modes are determined by  the time-inhomogenous stochastic events,
i.e.  the jump rates triggering changes in $E$ and $T$ depend on  $I(t)$\cite{NoiJTB}.
The exit times for mode $q$ are given by the time-dependent cumulative distribution function 
\begin{align}\label{eq:Jumptime}
\Probab_q[\tau]= \exp\left(\sum_{i}^{} \int_{0}^{\tau}{a_{i,q}(t_q+t)dt} \right)
\end{align}
and the probability of jumping to mode $q'$, given the exit time $\tau$, is
\begin{align}\label{eq:Jumpmode}
\Probab_q[q'\mid\tau]= 
\begin{cases}
\dfrac{\sum_{j \in Q}a_{j, q} (t_q+\tau)}{\sum_{i}^{}{a_{i, q}(t_q+\tau)}} & \text{ if }
Q= \{ j \mid q + \nu_j = q'\} \\
0 & \text{otherwise}\, .
\end{cases}
\end{align}
Notice that two stochastic events, i.e. $a_{2,q}$ and $a_{3,q}$, trigger jumps to the same new mode, i.e. jumps from $q=(q_E,q_T)$ to $(q_E-1,q_T)$, so  their probabilities  sum up in  $Q$.
Here the Gillespie-like \cite{G76} notation is used so  $\nu_j$ is the $j$-th column of the system {\em stoichiometry matrix}
$$
\nu =\left(
\begin{matrix}
1&-1&-1&0&0&0 \\
0&0&0&1&-1&1
\end{matrix}
\right) 
$$
and the jump rates  in mode $q=(q_T,q_E)$ are the time-dependet {\em propensity functions} \cite{G77}
\begin{align*}
a_{1, q} (t)&=  r_2 q_T  && a_{2, q}(t)= r_2 bV^{-1} q_T(q_T-1)  \\
a_{3, q}(t)&= (p_T q_T q_E)/(g_TV + q_T)  && a_{4, q}(t)=  [p_E q_E I(t)] / [g_E + I(t)]  \\
a_{5, q}(t)&=  \mu_E q_E && a_{6, q}(t)= cq_T  \, .
\end{align*}
Notice that all but $a_{4,q}$ are time-homogenous jump rates, i.e. do not depend on the $I(t)$ inside the mode, but, because of $a_{4,q}$ the underlying stochastic process is not homogenous.

\begin{figure}
\begin{center}$
\begin{array}{c} 
\includegraphics[width=5cm]{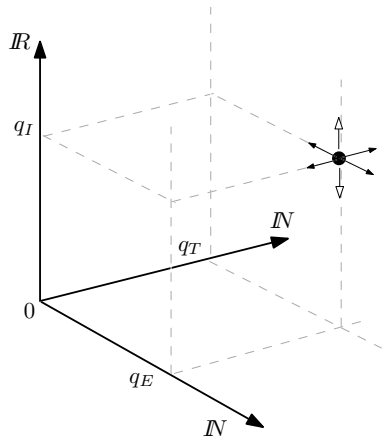}
\end{array}$
\end{center}
\caption{{\bf State space for the hybrid model.} The state space $\mathbb{N} \times \mathbb{N} \times \mathbb{R}$ for the PDMP \cite{PDMP} underlying the hybrid model when $\theta=0$. Once the process enters state $(q_T, q_E, q_I)$ the only  movement gradient is on the $z$-axis, i.e. the horizontal component $(q_T, q_E)$ is fixed and the process moves according to the vertical vector field represented by the empty arrows. The process persists moving  according to equation (\ref{eq:Jumptime}), and then moves on the $\mathbb{N} \times \mathbb{N}$ sub-space, i.e. the horizontal discrete grid denoted by the full arrows, according to equation (\ref{eq:Jumpmode}). When $\theta>0$ the process is enriched with a clock structure as for GSMPs \cite{GSMP1}, thus  inducing further jumps on the horizontal discrete grid to account for delayed reactions. }
\label{fig:state-space}
\end{figure}

Executions of this SHA are trajectories of the underlying {\em Piecewise Deterministic Markov Process} (PDMP, \cite{PDMP}), a jump process over vector fields which behaves deterministically  and whose jumps are triggered by $(i)$ hitting user-defined boundaries of the state space and $(ii)$ time-inhoumogenous jump distributions. Actually, for this case, the underlying PDMP has no hitting boundaries but only time-dependent jump rates linked to the vector field $I(t)$. The state space for the PDMP is $\mathbb{N} \times \mathbb{N} \times \mathbb{R}^+$, as shown in Figure  \ref{fig:state-space}. In there, once the process enters state $(q_T, q_E, q_I)$ the only  movement gradient is on the $z$-axis, i.e. the horizontal component $(q_T, q_E)$ is fixed and the process moves according to the vertical vector field. The process persists moving  according to equation (\ref{eq:Jumptime}), and then moves on the $\mathbb{N} \times \mathbb{N}$ sub-space, i.e. the horizontal discrete grid, according to equation (\ref{eq:Jumpmode}).

When $\theta > 0$ the SHA jumps are no more given by a continuous time Markov process but, instead, by  a {\em Generalized Semi-Markov Process} (GSMP, \cite{GSMP1}), a kind of process characterizing a large class of discrete-event simulations \cite{GSMP2,GSMP3,GSMP4} \footnote{
Theoretically, this process might  be equally reframed as a pure PDMP with unbounded number of clocks and  infinite dimensional state space.  Even though  proving existence and uniqueness of the solutions of the ODE would be feasible, we think that the combined process allows for the definition of an  efficient simulation algorithm (see Section \ref{sec:analysis}).}. It is shown in \cite{IoJane,Iotesi} that these process underly Gillespie-like\cite{G77} chemically reacting systems with deterministic delays, those indeed used here.  In these discrete  processes  $(i)$ the  embedded state process is a Markov chain and $(ii)$ the  time between jumps is an
 arbitrarily distributed  random variable which may  depend on the  starting and the ending modes.
When  $(a)$ a single  jump event is present in each state then the  process is a Semi-Markov Process,  when  $(b)$ multiple are currently running then the process is a GSMP and, finally,  when  $(c)$ the jump times are  exponentially distributed, i.e. memoryless, then the GSMP becomes a Continuous-Time Markov Chain (CTMC). 

We recall the definition of finite-state GSMPs as in \cite{GSMP2}; the overall process will have the structure of the PDMP  with the GSMP clock structure superimposed. We remark that, even if the state-space of our process is not finite, i.e. both $T$ and $E$ can theoretically grow unbounded, we could arbitrarily define two thresholds to limit the cells growth to account for biologically realistic configurations. Regions of the parameters in which unbounded growth of the cellular populations are determined in \cite{KP,NoiJTB}, and could be used to define such thresholds. Here, since we only perform simulation-based analysis of these processes we can avoid restricting the GSMP to a finite state space.
Let $\EE=\{e_1, \ldots, e_n\}$ 
be a finite set of {\em  events} and, for any state $s \in S$, let $s \mapsto E(s)$ be a mapping 
from $s$ to a non-empty subset of $\EE$ denoting the {\em active events} in  state $s$. 
In this GSMP one  exponential event is always the one related to the jump process, and there is one event for each delayed transition pending; in next section  an algorithm to simulate this joint process is given. When in state $s$ the occurrence of one or more events triggers a state transition, the next 
state $s'$  is chosen according to a probability distribution $p(s'; s, E^\ast)$ where 
$E^\ast \subseteq E(s)$ is the set of active events which are triggering the state transition. 
Clocks are associated with events and, in state $s$, the clock associated with event $e$ 
decays at rate $r(s,e)=1$ since, in this  case, time flows uniformly for the involved components.  When, in a state $s$, there are no outgoing transitions, i.e.\ $E(s) = \emptyset$, the state $s$ is said to be \emph{absorbing} and it models a terminating process. 
The set of possible {\em clock-reading vectors} when the state is $s$ is
\[ 
C(s) = \left\{
c=(c_1, \ldots, c_M) \; \mid\;
 c_i \in [0, \infty) \wedge 
 c_i >0 \Leftrightarrow e_i \in E(s)
\right\}
\] 
where $c_i$ is the value of the clock associated with $e_i$; $c_i \in \mathcal{C}_\ell$ where $\mathcal{C}_\ell$
is the set of clock evalutions. In state $s$ with clock-reading 
vector $c=(c_1, \ldots, c_M)$, the time to the next transition is 
\[
t^\ast(s, c) = \min_{\{i \mid e_i \in E(s)\}} c_i / r(s ,e_i) = \min_{\{i \mid e_i \in E(s)\}} c_i 
\]
where $ c_i / r(s ,e_i) = +\infty$ when $r(s ,e_i)=0$. The set of events triggering the state 
transition is then
\[
E^\ast (s,c) = \left\{
e_i \in E(s) \mid c_i - t^\ast(s,c)r(s, e_i) = 0
\right\} \, .
\]
Actually, as is shown in \cite{IoJane}, by probabilistic arguments it is possible to show that, for chemically reacting systems with delays, there is a unique possible events triggering at once, i.e. $E^\ast (s,c)$ is a singleton.
When a state transition from $s$ to $s'$ is triggered the events $E^\ast$ expire, leaving 
$E'(s) = E(s)\setminus E^\ast$.  Moreover some new events are created; this set of
\emph{new events} is $E(s') \setminus E'(s)$. For these events $e'$ a clock value $x$ is 
generated by a \emph{distribution-assignment} function $F(x; s', e', s, E^\ast)$ such that 
$F(0; s', e', s, E^\ast) = 0$ and $\lim_{x\to \infty} F(x; s', e', s, E^\ast) = 1$. For the {\em old events}
 in $E(s') \cap E'(s)$ the clock value in state $s$ at the time when the transition was triggered 
 is maintained in $s'$. In $s'$ events in $E'(s) \setminus E(s')$ are cancelled  and the 
 corresponding clock value is discarded. The GSMP is a continuous-time stochastic process 
 $\{X(t) \mid t \geq 0\}$ recording the state of the system as it evolves and its semantics is given 
 in terms of a general state space Markov chain storing both the state of the process and the 
 clock-reading vectors \cite{GSMP1}. 
 
%

\begin{algorithm}[t]
\caption{Input: $(T_0, E_0, I_0)$, start time $t_0$, stop time $t_{stop}$}
\label{alg:dssa-hyb-nodelay}
\begin{algorithmic}[1]
\STATE set  initial mode  $q \gets (q_{T_0}, q_{E_0})$ and set $I(t_0)=I_0$;
\WHILE{$t < t_{stop}$} 
\STATE let $r_1 \sim U[0,1]$ determine the mode exit time $\tau$ as $\Probab_q[\tau] = 1/r_1$  thus solving equation (\ref{eq:Jumptime});
\STATE determine the jump rates $a_{j,q}(t+\tau)$, set $I(t+\tau)$;
\STATE jump to mode $q'$ with probability $\Probab_q[q'\mid\tau]$;
\ENDWHILE
\end{algorithmic}
\end{algorithm}

%
%
%
%

\section{Simulating the model} \label{sec:analysis}
We present here  an algorithm to realize trajectories of the the underlying PDMP with the superimposed GSMP clock structure  and   provide model parameters.

\paragraph{Model simulation.} When $\theta=0$ the SHA trajectories are generated by  Algorithm \ref{alg:dssa-hyb-nodelay}, an extension of the Gillespie {\em Stochastic Simulation Algorithm} (SSA) \cite{G76,G77} accounting for time-dependent jump rates and specifically tailored for this hybrid system \cite{NoiBMC}. Jump times are given by solving  equation (\ref{eq:Jumptime}).

When $\theta>0$  a combination of such an algorithm  with the {\em SSA with Delays} (DSSA, \cite{Barrio,Iotesi}) is required. The DSSA  generates a statistically correct trajectory of the  GSMP underlying chemically-reacting systems with delays \cite{Iotesi,IoJane}. Practically, such an algorithm is  the SSA wrapped within an acceptance/rejection scheme  to schedule/handle reactions with delays.
Thus, the DSSA provides an algorithmic approach to the solution of the {\em Delay Chemical Master Equation} (DCME, \cite{Iotesi}), the non-Markovian master equation ruling chemically reacting systems with delays. In this hybrid case, the system master equation is defined over  the hybrid state-space \cite{Gardiner,Bounded} and extended to account for the delays, i.e. it is a {\em differential Chapman Kolmogorov} equation with delays.

\begin{algorithm}[t]
\caption{Input: $(T_0, E_0, I_0)$, start time $t_0$, stop time $t_{stop}$}
\label{alg:dssa-hyb}
\begin{algorithmic}[1]
\STATE set  initial mode  $q \gets (q_{T_0}, q_{E_0})$, set $I(t_0)=I_0$ and empty scheduling list ${\Pi}$;
\WHILE{$t < t_{stop}$} 
\STATE let $r_1 \sim U[0,1]$ determine the mode exit time $\tau$ as $\Probab_q[\tau] = 1/r_1$;
\IF {$\tau < \texttt{head}(\Pi)$}
%
\STATE determine the rate triggering the jump according to  $a_{j,q}(t+\tau)$, set $I(t+\tau)$;
\IF {the jump is triggered by $a_{6,q}$}
\STATE stay in mode $q$, set $t \gets t+\tau$ and schedule, i.e. $\texttt{enqueue}(t+\tau+\theta, Pi)$;
\ELSE
\STATE jump to mode $q'$ with probability $\Probab_q[q'\mid\tau]$;
\ENDIF
\ELSE
\STATE let $\tau'=\texttt{head}(\Pi)$,  jump to mode $(q_E + 1, q_T)$, set $I(t+\tau')$, $\texttt{dequeue}(\Pi)$ and $t\gets t+\tau'$;
\ENDIF
\ENDWHILE
\end{algorithmic}
\end{algorithm}

We present here Algorithm \ref{alg:dssa-hyb} which, at the best of our knowledge, is the first attempt to combine an algorithm for hybrid systems with delays, in the context of biological Gillepie-like systems. This should, in turn, suggest further extensions towards the formal definition of SHA with delays. The algorithm uses a acceptance/rejection  scheme and a scheduling list $\Pi$, as other DSSAs do. In this case, since a unique reaction with constant delay is present, $\Pi$ is a standard {\em queue data structure} offering \texttt{head}, \texttt{dequeue} and \texttt{enqueue}  operations. The algorithm works by determining, at each iteration, both the exit time from the current mode and the next mode, if any, or the scheduled reaction to handle. So, when at time $t_q$ the automaton enters a mode $q$, the exit time $\tau$  (step 3) is determined by the parallel solution of  $I(t)$, $t\geq t_q$, and  $\Probab_q[\tau]$ as 
triggered by the  jump rates $a_{j,q}(t)$. As in \cite{NoiJTB}, samples from   $\Probab_q[\tau]$  are obtained by a unit-rate Poisson transformation (step 3), i.e.
\[
\sum_{i}^{} \int_{0}^{\tau}{a_{i,q}(t_q+t)dt} =\ln\left(\dfrac{1}{r_1}\right)
\]
with $r_1$ uniformly distributed. Notice that in this equation, whose analytical solution is unknown, the computation is speeded up by using the analytical definition of $I(t)$, i.e. equation (\ref{eq:Imode}). 
If no reactions with delays are scheduled to complete in $[t_q, t_q+\tau]$, i.e. $\tau < \texttt{peek}(\Pi)$, the new mode is chosen as in the SHA for $\theta=0$ by a weighted probabilistic choice depending on $a_{j,q}(t+\tau)$, i.e.  the $j$ satisfying 
\[
\sum_{i=1}^{j-1} a_{i,q}(t+\tau) < r_2 \sum_{k=1}^6 a_{k,q}(t+\tau) \leq \sum_{i=1}^j  a_{i,q}(t+\tau)
\] 
with $r_2$ uniformly distributed.
However, if the jump is induced by the rate with delay, i.e. $a_{4,q}$, the automata stays in mode $q$  and the effectors recruitment is scheduled at time $t_q+\tau+\theta$ by means of the \texttt{enqueue} operation. This corresponds to assuming  the {\em purely delayed}  interpretation of delays \cite{Iotesi,PDA}, being a reaction with no reactants. Finally, if a reaction with delay is scheduled in  $[t_q, t_q+\tau]$, then the jump time is rejected, the system moves to the time at which the reaction is scheduled, a new effector cell is recruited, i.e. the system jumps from mode $(q_E, q_T)$  to mode $(q_E + 1, q_T)$ and the scheduled reaction is dequeued from $\Pi$.

\paragraph{Model parameters.}
We use   parameter values taken from \cite{NoiJTB}. The baseline growth rate of the tumor is  $r=0.18\, days^{-1}$ and the organism  carrying capacity is $b=1/10^{9}\, ml^{-1}$.
The  baseline strength of the killing rate of tumor cells by $E$, of the $IL-2$-stimulated growth rate of $E$ and of the production rate for $I$ are, respectively,  $p_T=1 \,ml/days$, $p_E=0.1245\, days^{-1}$ and $p_I=5\, pg/days$. The corresponding $50\%$ reduction factors are  $g_T=10^5\, ml^{-1}$, $g_E=2\cdot 10^7\, pg/l$  and  $g_I=10^3\, ml^{-1}$, respectively. The degradation rates are $\mu_E=0.03\, days^{-1}$ for the inverse of the average lifespan of $E$ and $\mu_I=10\, days^{-1}$ for the loss/degradation rate of $IL_2$. Finally,  the reference volume is  $V=3.2\, ml$.

These values pertain to mice \cite{KP,Arciero} and are taken from \cite{DeBoer,[Kuznetsov]}, where accurate fitting of real data concerning laboratory animals were performed. Volume $V$, instead, has been estimated in \cite{NoiJTB} by considering the body weight and blood volume of a chimeric mouse. The value of $\theta$ and $c$ are varied in each configuration and given in the   captions of figures.

\section{Results} \label{sec:results} 
With the purpose of investigating the effect of different delays on the tumor eradication time, if any, and on  the tumor  growth size, we performed extensive  simulations of various model configurations. 
All the simulations have been performed by  a \textsc{Java} implementation of the model running on the cluster \texttt{scilx.disco.unimib.it}, i.e. $15$ dual-core nodes, $2.0\, Ghz$ processors and $1\, GB$ of memory. Simulation times increase as $T$ and $E$ increase in size,  spanning from  few minutes to some hours, thus requiring a cluster capabilities to perform thousands of simulations in reasonable time.

\begin{figure}[t]
\begin{center}$
\begin{array}{cc} 
\includegraphics[width=7cm,height = 6 cm]{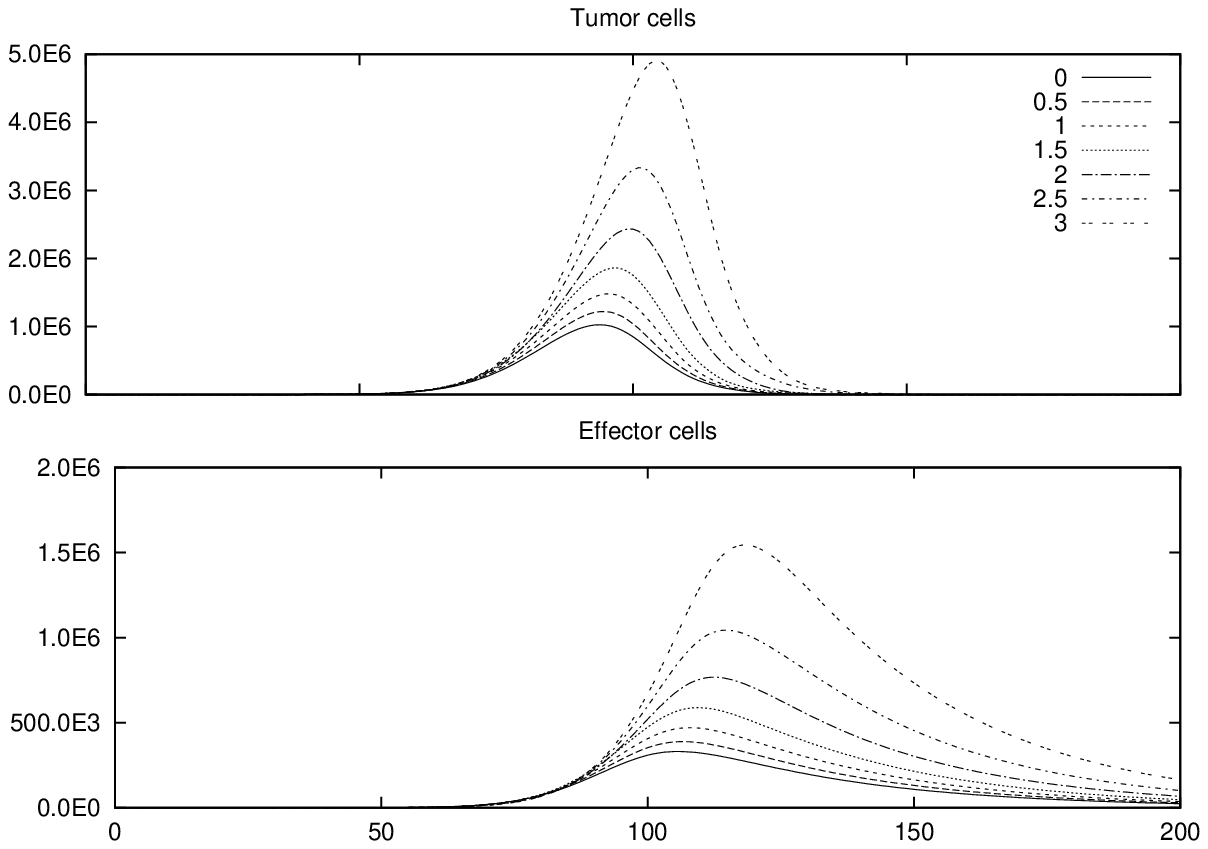} &
\includegraphics[width=9cm, height = 6 cm]{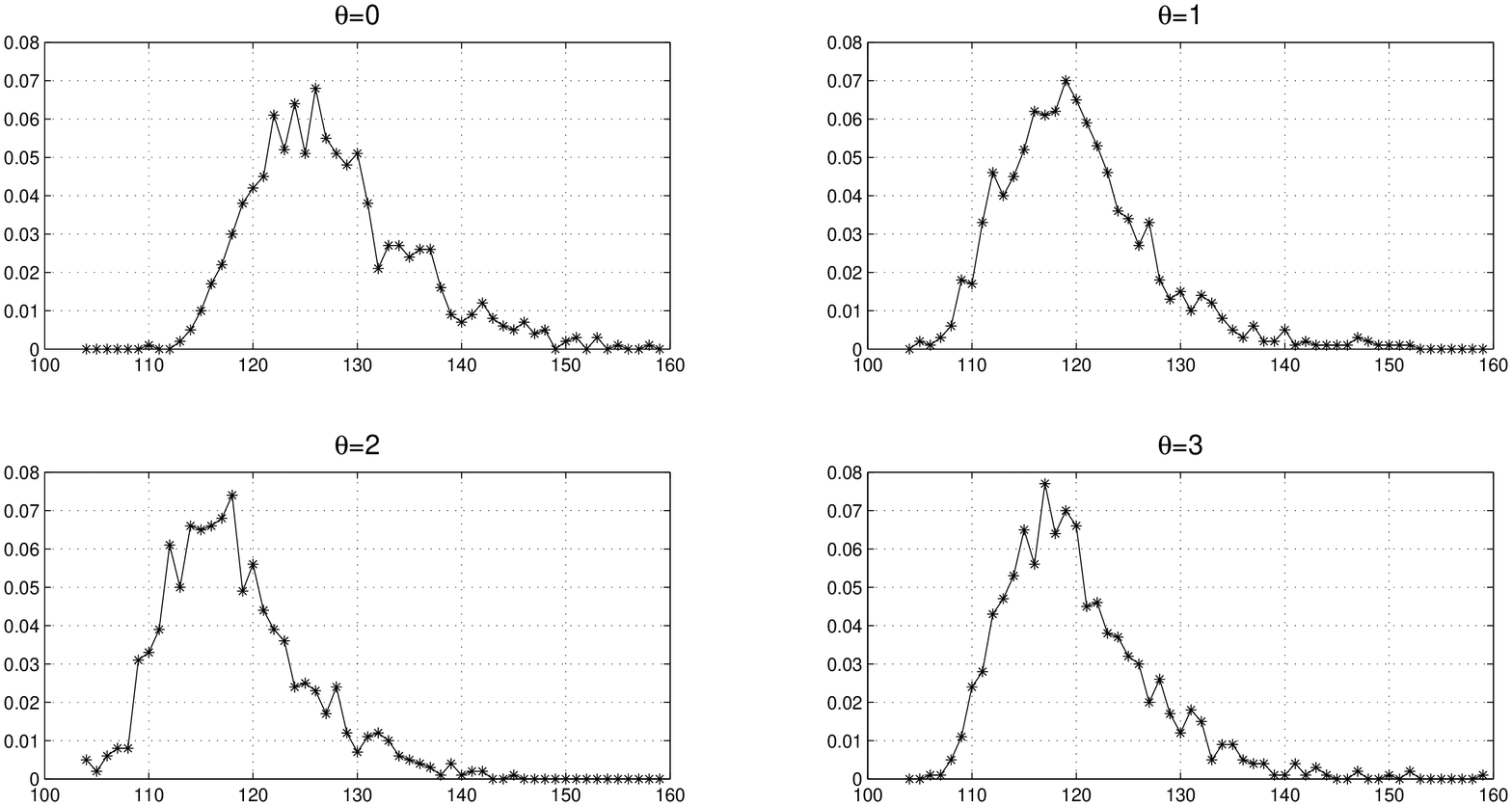} 
\end{array}$
\end{center}
\caption{{\bf Tumor and Effectors growth and eradication times.} In left we plot the average growth $\langle T(t)\rangle$ and $\langle E(t)\rangle$ as of  $10^3$ simulations with $c=0.02$,  $\theta \in \{0,  0.5, 1, 1.5, 2, 2.5, 3\}$ as reported in the legend and $(T_0, E_0, I_0)=(1,0,0)$. On the $x$-axis {days} are represented, on the $y$-axis number of cells. In right we plot the empirical probability density of the eradication time, i.e. $\Probab[T(t)=0]$ with $t\in \mathbb{N}$, for $\theta \in \{0, 1, 2, 3\}$. On the  $y$-axis  probability density are represented.}

\label{fig:c2E-2-averages}
\end{figure}

We always used the initial condition $(T_0, E_0, I_0)=(1,0,0)$, one of those used in \cite{NoiJTB} where also the effect of  an initial  bigger tumor or effectors mass is investigated.  However, we here use this initial condition since it allows to observe various qualitative behaviors \cite{NoiJTB}. For $c=0.02$, a value used in Figure 2 of \cite{NoiJTB}, we used $\theta \in \{0,  0.5, 1, 1.5, 2, 2.5, 3\}$ since, for $\theta > 3$, it is shown in \cite{DelayTumor} that the tumor mass grows up to the carrying capacity of the organism, i.e. $1/b$. We remark that $\theta$ units are {\em days}, and $\theta > 3$ is a  biologically unrealistic value as shown in \cite{DelayTumor}. We performed $10^3$ simulations for each delay configuration, and we plot in Figure \ref{fig:c2E-2-averages} the averages  tumor and effectors growth, i.e. $\langle T(t)\rangle$ and $\langle E(t)\rangle$. 

Notice that, even though in each configuration the model still predicts tumor eradication, the tumor mass grows significantly more for higher delay values, i.e. for $\theta=0$ it reaches around $10^6$ cells whereas  for $\theta=3$ it is $5$ times bigger. This, in turn, stimulates the immune-response as shown by the plots of  the empirical probability density of the eradication time, i.e. $\Probab[T(t)=0]$ with $t\in \mathbb{N}$. Notice that, even though the state with $T=0$ is not absorbing in the GSMP, i.e. further reactions would lead to the  natural death of effector cells ending to the absorbing $(0,0)$ state, this corresponds to estimating the expected time for a quasi-absorbing state.
These plots suggest that, though the tumor mass grows more and more rapidly for higher $\theta$ -- as one might expect -- the effect of the consequent immune response is also larger, inducing a quicker eradication of the tumor, given that the mean peak for $\theta=0$ is around day $125$, whereas for  $\theta\in \{1,2,3\}$ is around day $120$, $118$ and $115$, respectively. This is a rather counterintuitive result, which hints at a functional role of delay in controlling the expansion of the tumor mass.



In order to quantitatively determine the  sensitivity of   tumor  growth with respect to $\theta$, we perform {\em parametric sensitivity analysis} (PSA) by using the technique defined in  \cite{Damiani2012}, which we now briefly recall. This is a  numerical  procedure specifically defined for discrete stochastic models; it is numerical since   models are only rarely analytically solvable. As model output variable we use the whole $\Probab[T(t)]$  rather than, for instance,  its overall mean or mode,  to capture dramatic variations in  $\Probab[T(t)]$, potentially induced by small perturbations on $\theta$.  Besides, we scan a wide range of values for $\theta$, given that the overall dynamics can be differently sensitive  in various regions of the parameters space. 
For this reason, in \cite{Damiani2012} the model  sensitivity  to a given parameter is defined as a function of the parameter itself. Differently from the mean-field case, where just $T(t)$ could be used, the stochastic sensitivity  is computed  as in \cite{costanza} 
\begin{equation}
\label{costanza_coefficient}
s_{\theta}(t)=\frac{\partial \Probab_\theta[T(t)]}{\partial \theta}
\end{equation}
where $\Probab_\theta[T(t)]$ is the probability of the tumor mass, given a value of $\theta$. The sensitivity analysis is then  based on a  measure for discrete stochastic systems  or, analogously, for the  discrete part of hybrid systems, obeying a generic chemical master equation \cite{gunawan}, i.e.
\begin{equation}
\label{gunawan_coefficient}
S_{T}(t,\theta)=E \left [ | s_{\theta}(t) | \right ] = \int_{\mathbb{N}} \biggl| \frac{\partial \Probab_\theta[T(t)=x]}{\partial \theta} \biggl| \Probab_\theta[T(t)=x]dx \, .
\end{equation}
%
The dependency of $\Probab_\theta[T(t)]$ with respect to $\theta$ is then represented by a curve, 
which should be obtained as a  function of a possibly large range of values of $\theta$, instead of punctual perturbations. Here it  is  obtained by interpolating the points with a polynome of order $D-1$, where $D$ is the number of different values of delay.  The model overall sensitivity coefficient,  which does not depend on  $\theta$, is then 
\begin{equation}
\label{integral_coefficient}
S_{T}(t)= \int_{\Omega \theta} S_{T}(t,\theta) d\theta 
\end{equation}
where the finite domain $\Omega \theta$ for $\theta$  is used. Notice that, since  densities integrate to $1$, the sensitivity coefficients do not require to be normalized as is the case for mean-field models. Also, the integral on $\mathbb{N}$ is discrete, and can be therefore represented as a summation. 

\begin{figure}[t]
\begin{center}$
\begin{array}{cc} 
\includegraphics[width=9.5 cm, height = 5 cm]{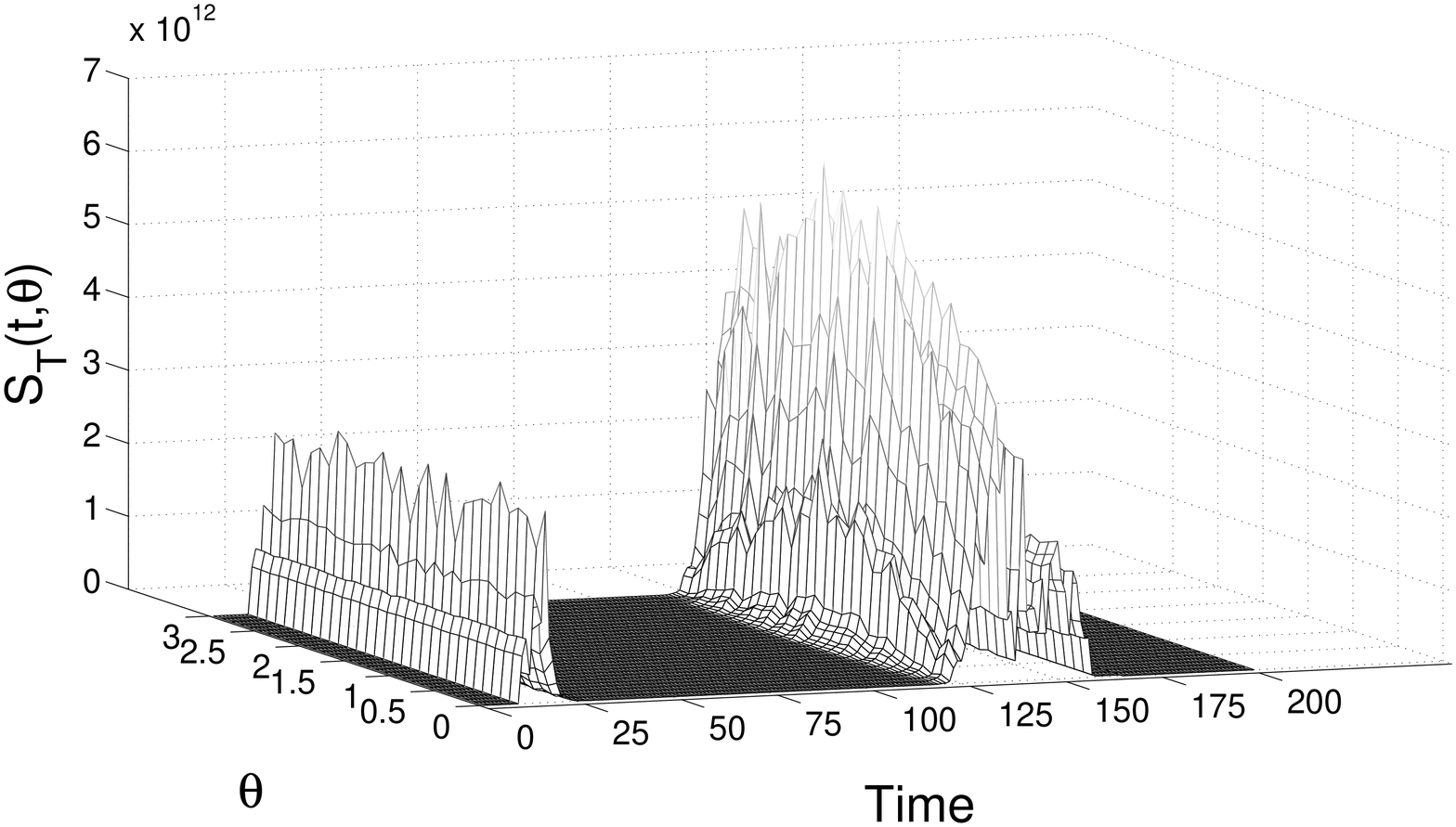} &
\includegraphics[width=6 cm, height = 4.7cm]{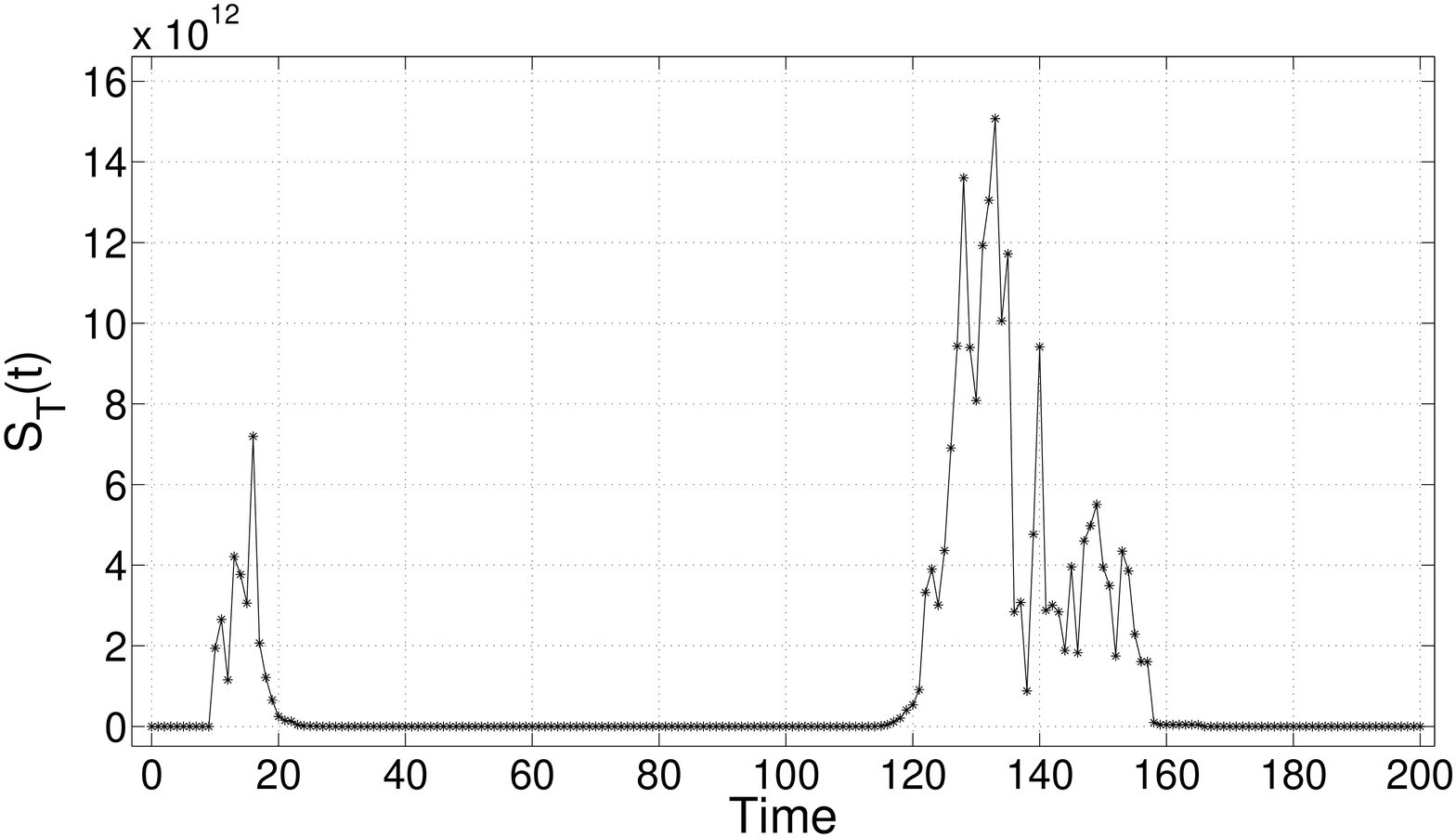} 
\end{array}$
\end{center}
\caption{{\bf Sensitivity analysis.} In left  we plot a 3-D representation of the sensitivity curves $S_{T}(t,\theta)$, plotted in correspondence of each delay value of the form $\theta=0.1\, k$ with $0\leq k \leq30$ and $k \in \mathbb{N}$, for each $t\in[0,200]$, i.e. equation (\ref{gunawan_coefficient}), as obtained by
$3\times 10^5$ independent simulations.  In right we plot the sensitivity curve $S_{T}(t)$ of equation (\ref{integral_coefficient}).}
\label{fig:C0.02-sensitivity_3d}
\end{figure}

To apply this technique  we performed $10^3$ simulations for each delay value in $\Omega\theta=\{0.1\, k \mid 0\leq k \leq30, k \in \mathbb{N}\}$, thus we use $3\times 10^5$ independent simulations, $D=30$ and every density function is computed on the range  $ [0; max_T]$, where $max_{T}$ is the maximum observed value of $T$ for all the values of $\theta$, in all the simulations. 
The sensitivity function $S_{T,\theta}(\theta,t)$ is then derived by integrating, for any  $\theta$, the absolute value of  the derivative $\partial \Probab_\theta[T(t)]/\partial \theta$ is evaluated in $x \in \mathbb{N}$ and weighted by $\Probab_\theta[T(t)=x]$ according to equation (\ref{gunawan_coefficient}). Notice that this method does not discriminate the sign of the observed variation\footnote{To perform PSA we  only adopted Lagrange polynomial interpolation, even though  multiple  interpolation methods could be used and compared, e.g. spline or other non-linear interpolation techniques.}. The  sensitivity curves, i.e.  equation (\ref{gunawan_coefficient}) and  (\ref{integral_coefficient}) are shown in Figure \ref{fig:C0.02-sensitivity_3d}.
%
%
%

One important general result is that the model sensitivity   to the variation of $\theta$ is not time-invariant, as shown in Figure \ref{fig:C0.02-sensitivity_3d}. It is indeed possible to detect two intervals in which the influence is maximum, i.e.   the intervals $[10,  25]$ and $[115, 160]$, while in the other regions the sensitivity is essentially not relevant. In particular, the overall sensitivity magnitude   is much larger in $[115, 160]$, almost doubling the  overall maximum of the first interval (right figure). 
This result suggests that a variation in the response time of the immune system can indeed influences the development of the tumor mass, but only in two specific conditions: $(i)$ before that the tumor begins its expansion (i.e. first interval), either preventing or favoring it; $(ii)$ after that the tumor has reached its maximum size, inducing either an enlargement or a reduction of the final eradication time. 
By looking at  $S_{T,\theta}(\theta,t)$ (left figure) it is then possible to notice that $(iii)$ in regard to the first interval,  the overall sensitivity is scarcely correlated to the specific  $\theta$, while $(iv)$ the sensitivity curves corresponding to $[115, 160]$ usually present a bell-shape, often characterized by a unique maximum value of sensitivity, with respect to a specific $\theta$. This  suggests that a variation in $\theta$  can provoke different repercussions on the overall dynamics in distinct regions of the parameter's space.

\begin{figure}[t]
\begin{center}$
\begin{array}{cc} 
\includegraphics[width=7.4cm]{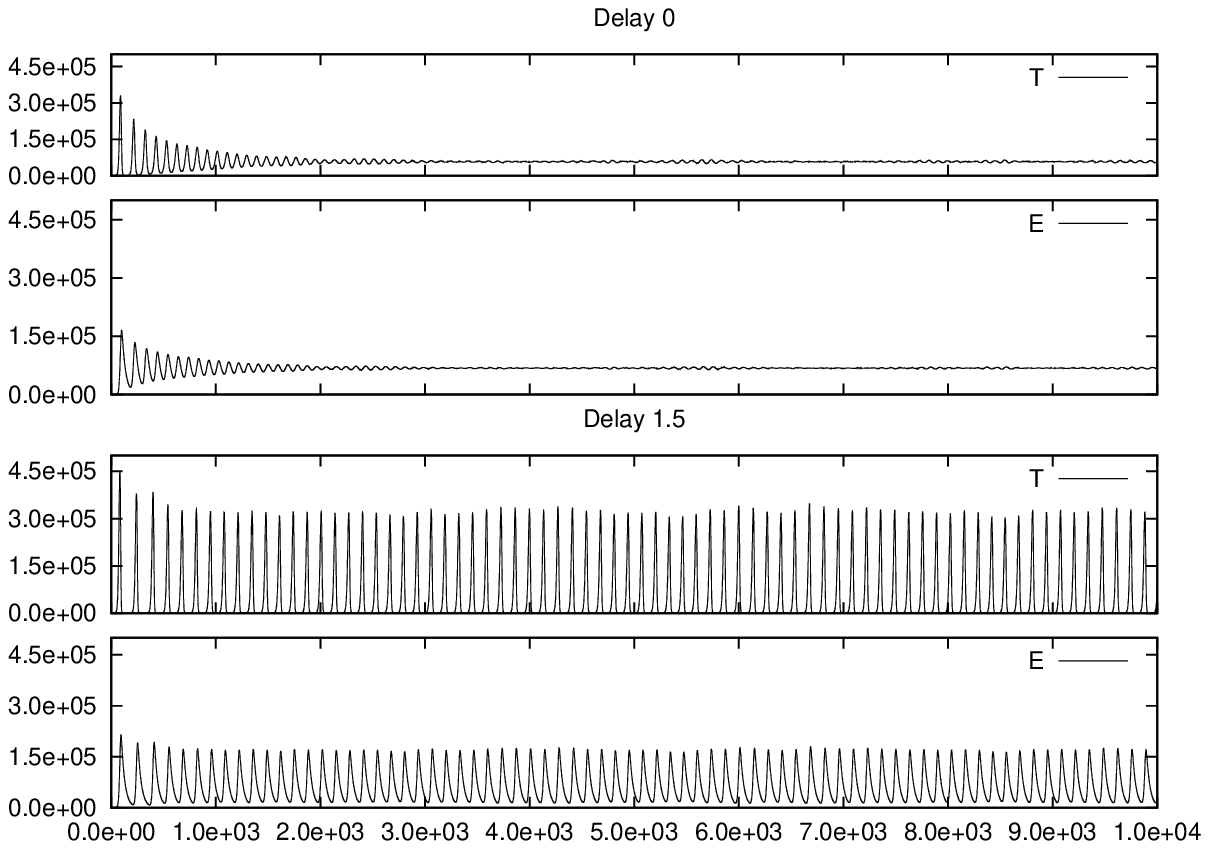} & 
\includegraphics[width=7cm]{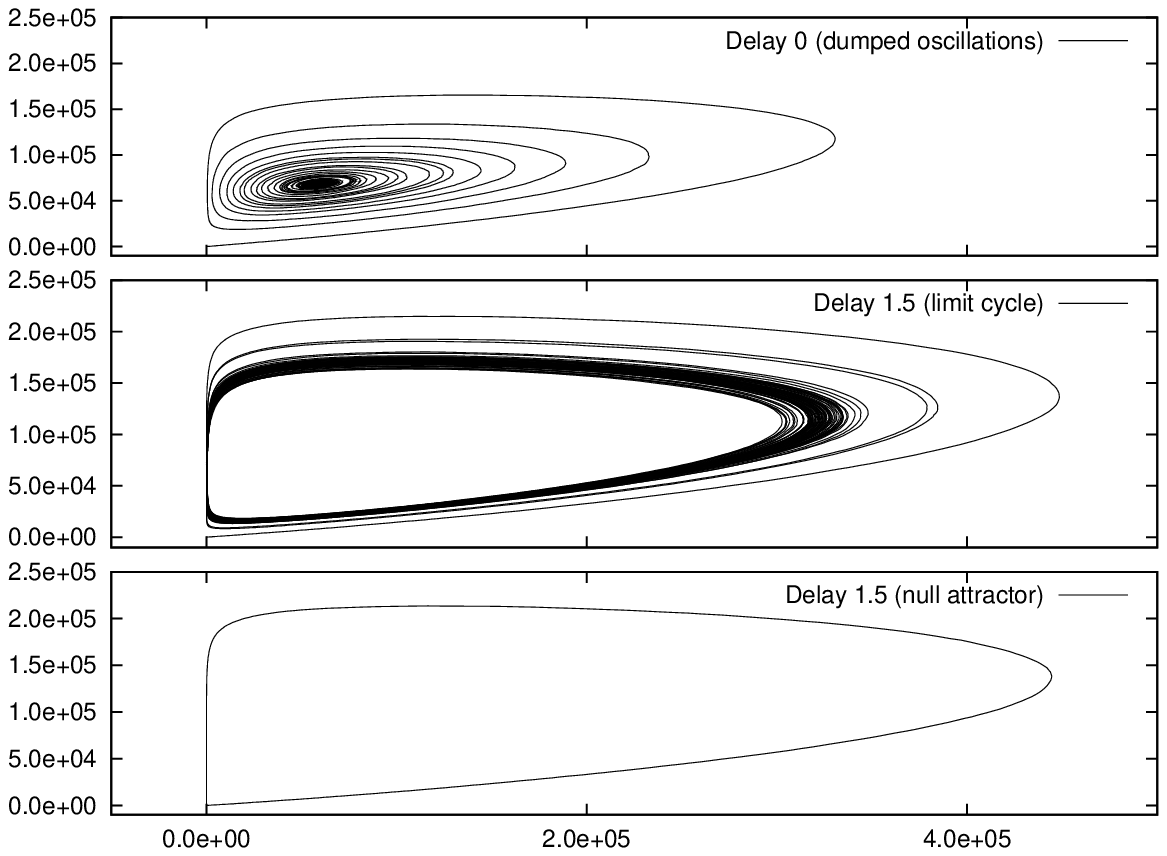}
\end{array}$
\end{center}
\caption{{\bf Stable oscillatory equilibria.} In left we plot $T(t)$ and $E(t)$ for a single run with $c=0.035$ and $\theta \in \{0,1.5\}$ as reported in the legend. The initial configuration is $(T_0, E_0, I_0)=(1,0,0)$. On the $x$-axis {days} are represented, on the $y$-axis number of cells. In right we plot the phase space of the system restricted to $T$ and $E$, and we show a stochastic switch to the null attractor for $\theta=1.5$.
}
\label{fig:C0.035-srun}
\end{figure}

In order to investigate the role of delays for the system in the oscillatory regime, we performed simulations with $0.03 \leq c \leq 0.035$, a region for which both the deterministic system   (i.e. Figure 2D of \cite{KP}) and the therapy-free hybrid model  (i.e. Figure 7 of \cite{NoiJTB}) predict tumor sustained/dumped oscillations.  In Figure \ref{fig:C0.035-srun} (left) we plot the effect of delays in the oscillatory regime for $c=0.035$,  $\theta \in \{0,1.5\}$ and initial configuration $(T_0, E_0, I_0)=(1,0,0)$. Here we simulate the model for around $10000$ days, i.e. $27$ years, a value far beyond the life expectancy of a mouse -- on which parameters are fitted -- but which serves mainly to prove the stability of the equilibrium, if any.

\begin{figure}[t]
\begin{center}$
\begin{array}{c} 
\includegraphics[width=7cm]{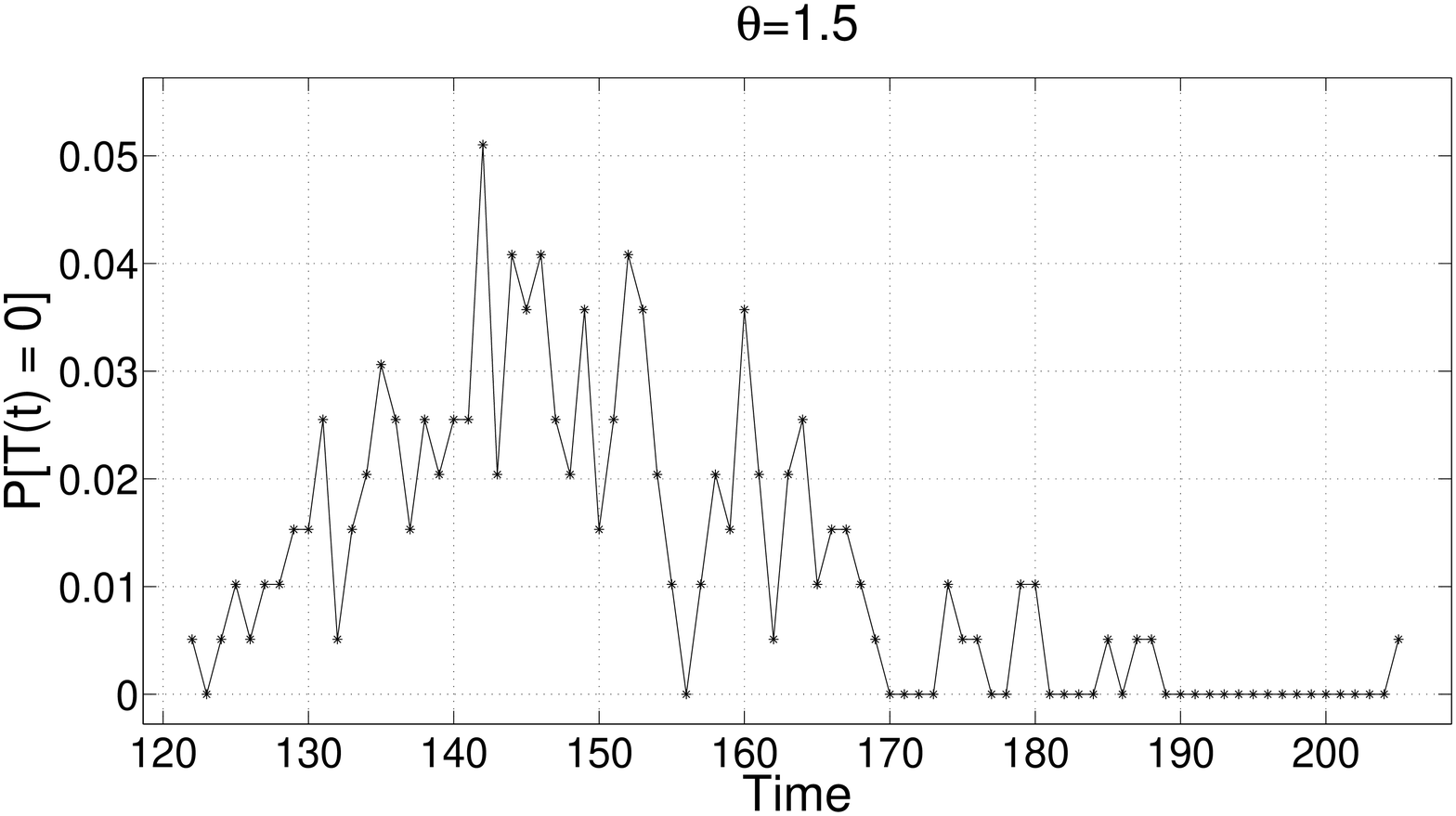} 
\end{array}$
\end{center}
\caption{{\bf Heuristic stochastic bifurcation for $\theta=1.5$.} We plot the empirical probability density of the eradication time, i.e. $\Probab[T(t)=0]$ with $t\in \mathbb{N}$, for  $c=0.035$, $\theta =1.5$ and $(T_0, E_0, I_0)=(1,0,0)$ as evaluated by the $196$ cases, out of $1000$, in which the system jumps to the null attractor for $T$.}
\label{fig:C0.035-pdf}
\end{figure}

It is immediate to notice that, for $\theta=1.5$ the tumor mass does not seem to reach a small equilibrium, as instead it happens for the delay-free case. Indeed, in the former case the tumor mass spans between very low values and  $3\times 10^5$, in the latter the oscillations are dumped up to around $10^5$ cells. Furthermore, the first oscillation peak is around $4.5\times 10^5$ for $\theta=1.5$ which is a considerably bigger values than that one reached for $\theta=0$. These  amplified oscillations often arise when models are enriched with delays \cite{Alb36,Alb37,Alb38} and reach very small values as shown in Figure \ref{fig:C0.035-srun} (right) where  the phase space of the system restricted to $T$ and $E$ is represented and a stochastic switch to the null attractor for $\theta=1.5$ is shown. Surprisingly, this result in some simulations showing eradication for $\theta=1.5$, an unexpected outcome for the oscillatory regime since for $\theta=0$  none of $1000$  simulations have  shown eradication (not shown here). Instead, $196$ out of $1000$ simulations, i.e. almost $20\, \%$ of the cases, for $\theta=1.5$ show eradication reached immediately after the first spike of the oscillations. This clearly suggests the existence of a heuristic stochastic bifurcation  close to $\theta =1.5$ with a switch to the null attractor for $T$, i.e. $T\to 0$, so that, for some cases, the tumor gets eradicated. In Figure \ref{fig:C0.035-pdf} we plot the empirical  probability density of the eradication time, i.e. $\Probab[T(t)=0]$ with $t\in \mathbb{N}$,
as evaluated by these $196$ cases. This conclusion is strengthened by observing that, for $\theta\in\{2,3\}$, the tumor is always eradicated (in $1000$ cases, not shown). 

\begin{figure}[t]
\begin{center}$
\begin{array}{cc} 
\includegraphics[width=7cm]{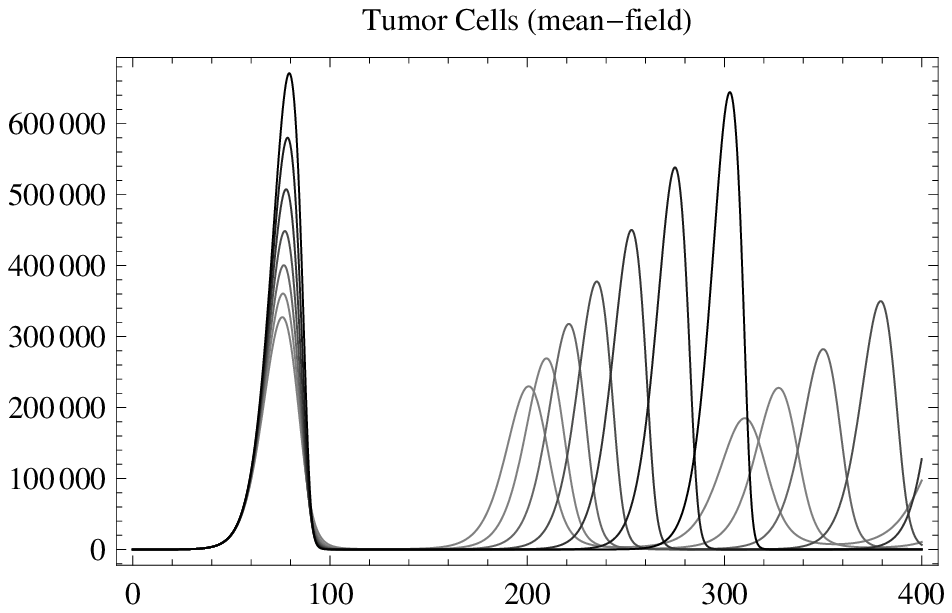} & 
\includegraphics[width=6.5cm]{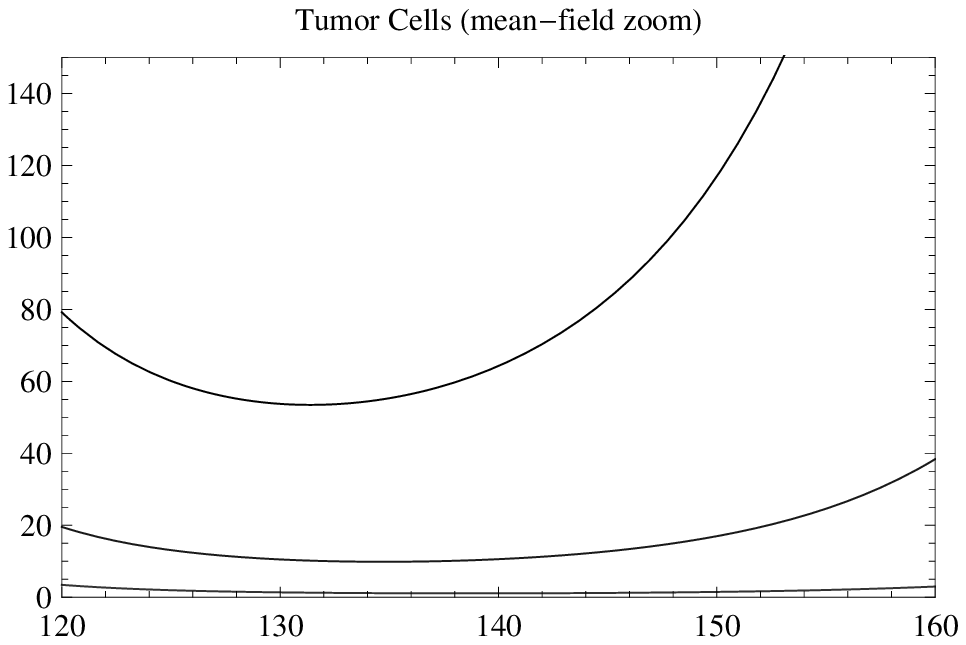}
\end{array}$
\end{center}
\caption{{\bf Mean-field model.} We plot deterministic simulations of model (\ref{eq:KPcells}-\ref{eq:KPil}) for  $c=0.035$,  $\theta \in \{0,  0.5, 1, 1.5, 2, 2.5, 3\}$ (higher peaks for higher values of $\theta$) and $(T_0, E_0, I_0)=(1,0,0)$, $T(t)=0$ for $t<0$. Notice  the tumor resting period in $t\in [120,160]$ (right zoom for $\theta \in\{1,1.5,2\}$), the length of which depends on $\theta$, is the one in witch the hybrid system probabilistically switches to the null attractor for $T$. On the $x$-axis {days} are represented, on the $y$-axis number of cells.
}
\label{fig:C0.035-dde}
\end{figure}

Moreover, this is an interesting outcome as compared against the predictions of the mean-field model. In fact, in Figure  \ref{fig:C0.035-dde} we show  deterministic simulations of model (\ref{eq:KPcells}-\ref{eq:KPil}) for $\theta \in \{0,  0.5, 1, 1.5, 2, 2.5, 3\}$, restricted to $t\in[0, 400]$ and with extended  analogous initial condition
\[
T(t) = \begin{cases} 0&  t<0 \\ 1 & t=0 \end{cases}, \qquad\qquad E_0=I_0=0 \,.
\]
 In there it is possible to observe a  tumor resting period for $t\in[120, 160]$, the length of which depends on $\theta$. Small values in such period are predicted, i.e. for $\theta =2$ we observe $T(t)<1$ and for  $\theta =1.5$ we observe $T(t)\approx 10$ in accordance with the simulations we performed.
 In this same period, instead, the hybrid system probabilistically switches to the null attractor for $T$, thus suggesting the importance of resetting the model in the hybrid setting which, as in \cite{NoiJTB}, is again proved to be more informative.

\section{Conclusions} \label{sec:conclusions}
In this paper we study the effect of a constant time delay in effectors recruitment in a tumor--immune system interplay hybrid model. The model, analogous of a well-known mean-field model \cite{KP}, was  proved to be more informative to forecast onco-suppression
by the immune system \cite{NoiJTB} as a  conjunction of the intrinsic
tendency of the immune system to oscillate,  significantly
evidenced by the deterministic model, with the intrinsic noise. This phenomen, which is triggered by the appearance of specific  neo-antigens resulting from genetic and epigenetic  events  characterizing tumor cells \cite{PA03}, is fundamental to the  immune surveillance hypothesis,  a promising approach to the treatment of cancer \cite{Dunn2004}.

Modeling such an interplay requires  considering  biological entities at multiple  scales. 
As such, tumor growth is an ideal object of hybrid modeling \cite{NoiJTB}. Extending the model  in \cite{NoiJTB} with delays allows to account that,  
due to both chemical transportation and  cellular differentiation/division, the influence of tumor on  effectors recruitment and proliferation  exhibits a lag period. Of course, an explicit model of the missing dynamical components, e.g. chemical signals, maturation and activation of T-lymphocytes, would be desirable but is currently unfeasible, also because of the lack of systematic data \cite{Todorovic}. 

In this paper we contextualized this model within Stochastic Hybrid Automata, when the delay is  $0$, so to give it a semantics in terms of Piecewise Deterministic Markov Processes \cite{PDMP}. When delays are present we combine the underlying  process with a clock structure for a  Generalized Semi-Markov process \cite{GSMP1}, as for  chemically reacting systems with delays \cite{Iotesi}. We present a novel algorithm to simulate this extended hybrid model and, via numerical analyses,  we quantitatively determined the effects of various delays on tumor mass growth and  determine the eradication times as probability distributions, under various configurations. Under these configurations we adopted a parametric sensitivity analysis technique to relate the tumor growth to the delay amplitude. Also, we have shown that the stochastic effects driving the system to the eradication can unexpectedly appear  even in the oscillatory regime. In fact, in there we proved the existence of a heuristic stochastic bifurcation, which is neither predicted by the mean-field model nor by the hybrid non-delayed model.   Thus,  despite our model   being a highly macroscopical and simplistic representation of the  tumor--immune system interplay, we have shown that it can provide useful insights on the multitude of possible outcomes of this  very fundamental and complex interaction, e.g. neoplasm evasion from immune control, 
immune surveillance and (dumped) oscillations.

As far as future works are concerned, a further combination of this model with the immunotherapies studied in \cite{NoiBMC} would be interesting. Also, the model itself could be extended so, for instance, the linear antigenic effect $c T(t-\theta)$ due to the tumor size could be corrected by assuming a delayed saturating stimulation. Similarly,  the assumption that $E^{\prime}$ linearly depends on $E$ could be corrected, as there are cases where this dependence might be non-linear, as outlined in \cite{DON1}. Moreover, more complex form of delays could be considered, along the line of those used in mean-field models \cite{DelayTumor}, e.g. weak/strong kernels. Finally, the mathematical formalization of hybrid automata with delays seems  missing, thus suggesting possible extensions to the hybrid automata theory, along with their analysis techniques.

\providecommand{\urlalt}[2]{\href{#1}{#2}}
\providecommand{\doi}[1]{doi:\urlalt{http://dx.doi.org/#1}{#1}}

\nocite{*}
\bibliographystyle{eptcs}

\end{document}